# LOW-LEVEL RADIOFREQUENCY SYSTEM UPGRADE FOR THE DALIAN COHERENT LIGHT SOURCE*

H.L. Ding[1][†], J.F. Zhu[‡], H.K. Li[1], J.W. Han, X.W. Dai, J.Y. Yang[1], W.Q. Zhang[1]
Institute of Advanced Science Facilities, Shenzhen, China
[1]also at Dalian Institute of Chemical Physics, Chinese Academy of Sciences, Dalian, China

*Abstract*

DCLS (Dalian Coherent Light Source) is an FEL (Free-Electron Laser) user facility at EUV (Extreme Ultraviolet). The primary accelerator of DCLS operates at a repetition rate of 20 Hz, and the beam is divided at the end of the linear accelerator through Kicker to make two 10 Hz beamlines work simultaneously. In the past year, we have completed the upgrade of the DCLS LLRF (Low-Level Radiofrequency) system, including setting the microwave amplitude and phase for two beamlines based on event timing, optimizing the microwave stability, and generating microwave excitation with the arbitrary shape of amplitude and phase. We added two special event codes and a repetition rate division of 10 Hz in the event timing system and set the microwave amplitude and phase by judging the event code in LLRF. The amplitude and phase stability of the microwave was improved with an intra-pulse feedforward algorithm. In addition, we have also generated microwave excitation with arbitrary amplitude and phase shapes to meet the dual beam operation in the future. Detailed information on functions or algorithms will be presented in this paper.

## INTRODUCTION TO THE DCLS

Dalian Coherent Light Source (DCLS) is currently the only FEL user device operating in the EUV in the world, playing an essential role in promoting various disciplines such as energy chemistry and material biology in China [1-3]. Figure 1 shows the building of the DCLS.

As presented in Fig. 2 and Table 1, the length of the DCLS is 150 m. It has deployed user experiments since 2017. DCLS operates in room temperature pulse mode, with a maximum repetition rate of 50 Hz and a laser pulse energy higher than 100 μJ. Its wavelength is 50-150 nm, continuously adjustable, and completely coherent.

Figure 3 shows the model diagram of the DCLS device. It mainly consists of a photocathode injector, a linear accelerator, a seed laser, undulators, two beamlines, and four experimental stations.

We have completed upgrading the DCLS LLRF (Low-Level Radiofrequency) system in the past year. The following sections will show the details.

___________________
* Work supported by the National Natural Science Foundation of China (Grant No. 22288201), the Scientific Instrument Developing Project of the Chinese Academy of Sciences (Grant No. GJJSTD20220001), and the Shenzhen Science and Technology Program (Grant No. RCBS20221008093247072).
† dinghongli@dicp.ac.cn
‡ zhujinfu@mail.iasf.ac.cn

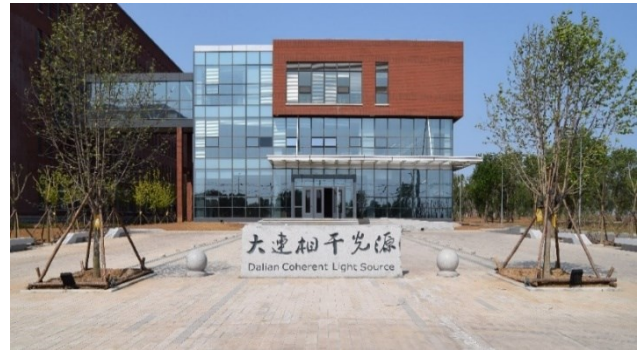

Figure 1: The building of the Dalian Coherent Light Source.

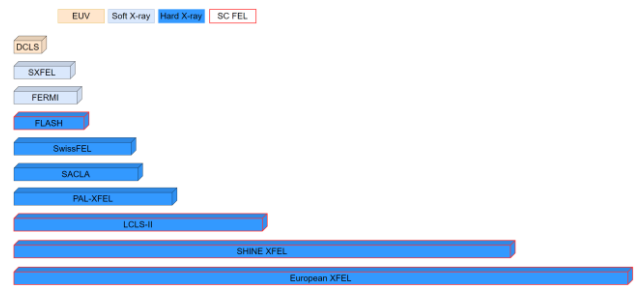

Figure 2: The DCLS and other FEL facilities.

## LLRF UPGRADE FOR TWO BEAMLINE OPERATIONS

As described in Fig. 4, the DCLS main oscillator (2856MHz) provides a low-jitter synchronous reference signal for the microwave system through the synchronization system. At the same time, the equipment of the microwave system needs a timing system to provide corresponding triggering timing. The DCLS microwave system consists of four pulse low-level devices (LLRF 1 ~ 4), each of which outputs excitation to drive four Solid-State Amplifiers (SSA). Four SSAs (SSA 1 ~ 4) drive two 50 MW klystrons (klystron 1 ~ 2) and two 80 MW klystrons (klystron 3 ~ 4), respectively. Each klystron is equipped with a high-voltage modulator (HV mod.). The high-power microwave output from the klystron is fed into the electron gun (GUN) and 7 S-band accelerating tubes through a high-power transmission system. Among them, the output power of the klystron one is provided to the electron gun and A0 accelerator tubes and klystron 2 ~ 4 for A1 ~ A6 accelerator tubes, respectively. The beam energy in the end reaches 300 MeV and enters the oscillator for modulation and FEL output.

Table 1: Specifications Comparison

| Country | Length / m | Electron beam energy / GeV | Wavelength range / nm | Repetition rate / Hz | Completion year |
|---|---|---|---|---|---|
| China | 150 | 0.3 | 50-150 | 50 | 2017 |
| China | 300 | 1 | 2-12.4 | 50 | 2020 |
| Italy | 350 | 1.2 | 4-40 | 10 | 2011 |
| Germany | 400 | 1.2 | 4-90 | 10 | 2005 |
| Switzerland | 715 | 5.8 | 0.1-7 | 100 | 2016 |
| Japan | 750 | 8 | 0.06-0.3 | 60 | 2011 |
| South Korea | 1100 | 10 | 0.06-10 | 60 | 2016 |
| U.S. | 1500 | 4 | 0.25-5 | 1,000,000 | 2023 |
| China | 3100 | 8 | 0.05-3 | 1,000,000 | Expected 2025 |
| Germany | 3800 | 17.5 | 0.05-4.7 | 27,000 | 2017 |

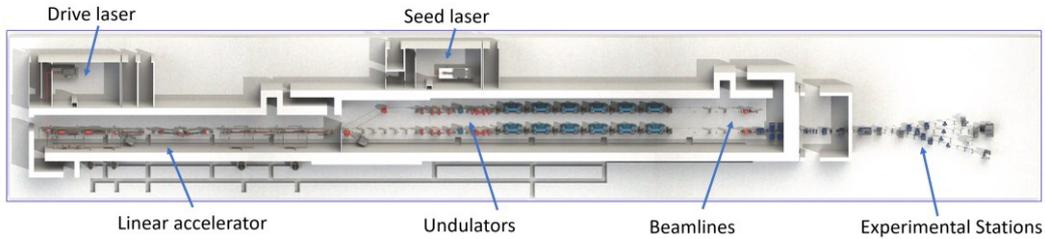

Figure 3: The model diagram of the DCLS device.

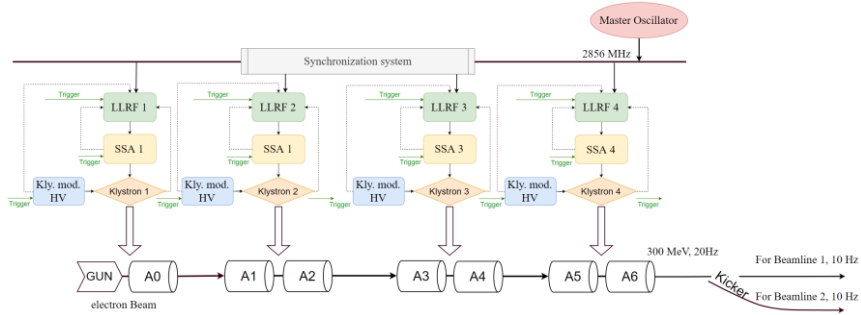

Figure 4: The microwave system of the DCLS.

The primary accelerator of DCLS operates at a repetition rate of 20 Hz now, and the beam is divided at the end of the linear accelerator through Kicker to make two 10 Hz beamlines work simultaneously. For the simultaneous emission FEL of two beamlines, the beam energy of the two beamlines is required to be controlled independently, so we need to set the amplitude and phase of each beamline's microwave. The user configuration interface is displayed in Fig. 5.

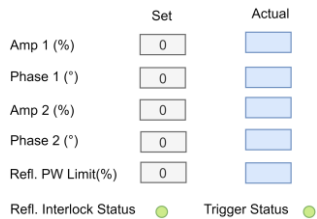

Figure 5: The user configuration interface for two beamline operations.

## IMPROVEMENT OF THE MICROWAVE STABILITY

During the pulsed operation of the linear accelerator in DCLS, we found a strong correlation between the klystron modulator's high voltage and the klystron output microwave, with noticeable jitter among adjacent microwaves. Therefore, we propose an intra-pulse feedforward algorithm and implement it in the LLRF. As indicated in Fig. 6, experiments have shown that this algorithm can effectively suppress the jitter among adjacent microwaves, e.g., improving the amplitude and phase stability (RMS) from 0.11%/0.2° to 0.10%/0.05°.

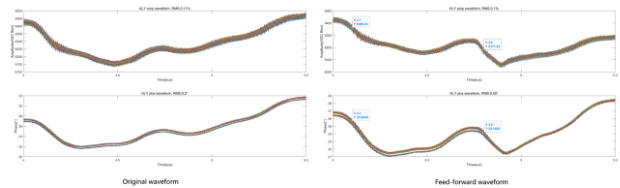

Figure 6: The comparison before and after feedforward.

# LLRF UPGRADE FOR THE DUAL BEAM OPERATION

We implement a microwave excitation whose amplitude and phase have arbitrary shapes in the LLRF system. For dual beam operation in DCLS, specifically, we generate a microwave pulse with step-shaped amplitude and phase. Due to the limited bandwidth of the klystron, the burst phase or amplitude of the step-shaped pulse may cause the excitation ring, so we generate a microwave excitation with a smooth step-shaped pulse by a sigmoid function. Figure 7 shows the amplitude waveform of klystron 1 forward and the GUN pick up.

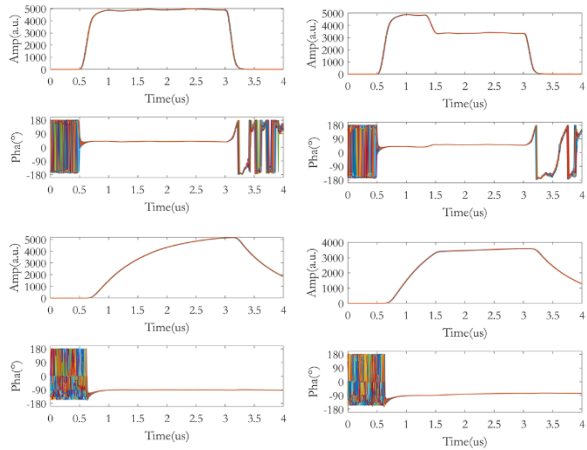

Figure 7: The microwave pulse amplitude and phase with step-shaped.

# CONCLUSION

This paper has a brief introduction to the DCLS LLRF upgrade. It involves upgrading LLRF firmware, software, and user interfaces for further operation.

# ACKNOWLEDGEMENTS

The authors would like to thank the Institute of High Energy Physics, Shanghai Advanced Research Institute, DESY, etc. who collaborated with the DCLS for their support and various discussions over the years.